\documentclass[astrosymb, preprint, modern]{aastex631}

\begin{document}

\title{Search for astrophysical electron antineutrinos in Super-Kamiokande
\\with 0.01wt\% gadolinium-loaded water}

\newcommand{\AFFicrr}{\affiliation{Kamioka Observatory, Institute for Cosmic Ray Research, University of Tokyo, Kamioka, Gifu 506-1205, Japan}}
\newcommand{\AFFkashiwa}{\affiliation{Research Center for Cosmic Neutrinos, Institute for Cosmic Ray Research, University of Tokyo, Kashiwa, Chiba 277-8582, Japan}}
\newcommand{\AFFicrronly}{\affiliation{Institute for Cosmic Ray Research, University of Tokyo, Kashiwa, Chiba 277-8582, Japan}}
\newcommand{\AFFipmu}{\affiliation{Kavli Institute for the Physics and
Mathematics of the Universe (WPI), The University of Tokyo Institutes for Advanced Study,
University of Tokyo, Kashiwa, Chiba 277-8583, Japan }}
\newcommand{\AFFmad}{\affiliation{Department of Theoretical Physics, University Autonoma Madrid, 28049 Madrid, Spain}}
\newcommand{\AFFubc}{\affiliation{Department of Physics and Astronomy, University of British Columbia, Vancouver, BC, V6T1Z4, Canada}}
\newcommand{\AFFbu}{\affiliation{Department of Physics, Boston University, Boston, MA 02215, USA}}
\newcommand{\AFFuci}{\affiliation{Department of Physics and Astronomy, University of California, Irvine, Irvine, CA 92697-4575, USA }}
\newcommand{\AFFcsu}{\affiliation{Department of Physics, California State University, Dominguez Hills, Carson, CA 90747, USA}}
\newcommand{\AFFcnm}{\affiliation{Institute for Universe and Elementary Particles, Chonnam National University, Gwangju 61186, Korea}}
\newcommand{\AFFduke}{\affiliation{Department of Physics, Duke University, Durham NC 27708, USA}}
\newcommand{\AFFfukuoka}{\affiliation{Junior College, Fukuoka Institute of Technology, Fukuoka, Fukuoka 811-0295, Japan}}
\newcommand{\AFFgifu}{\affiliation{Department of Physics, Gifu University, Gifu, Gifu 501-1193, Japan}}
\newcommand{\AFFgist}{\affiliation{GIST College, Gwangju Institute of Science and Technology, Gwangju 500-712, Korea}}
\newcommand{\AFFuh}{\affiliation{Department of Physics and Astronomy, University of Hawaii, Honolulu, HI 96822, USA}}
\newcommand{\AFFicl}{\affiliation{Department of Physics, Imperial College London , London, SW7 2AZ, United Kingdom }}
\newcommand{\AFFkek}{\affiliation{High Energy Accelerator Research Organization (KEK), Tsukuba, Ibaraki 305-0801, Japan }}
\newcommand{\AFFkobe}{\affiliation{Department of Physics, Kobe University, Kobe, Hyogo 657-8501, Japan}}
\newcommand{\AFFkyoto}{\affiliation{Department of Physics, Kyoto University, Kyoto, Kyoto 606-8502, Japan}}
\newcommand{\AFFliv}{\affiliation{Department of Physics, University of Liverpool, Liverpool, L69 7ZE, United Kingdom}}
\newcommand{\AFFmiyagi}{\affiliation{Department of Physics, Miyagi University of Education, Sendai, Miyagi 980-0845, Japan}}
\newcommand{\AFFnagoya}{\affiliation{Institute for Space-Earth Environmental Research, Nagoya University, Nagoya, Aichi 464-8602, Japan}}
\newcommand{\AFFkmi}{\affiliation{Kobayashi-Maskawa Institute for the Origin of Particles and the Universe, Nagoya University, Nagoya, Aichi 464-8602, Japan}}
\newcommand{\AFFpol}{\affiliation{National Centre For Nuclear Research, 02-093 Warsaw, Poland}}
\newcommand{\AFFsuny}{\affiliation{Department of Physics and Astronomy, State University of New York at Stony Brook, NY 11794-3800, USA}}
\newcommand{\AFFokayama}{\affiliation{Department of Physics, Okayama University, Okayama, Okayama 700-8530, Japan }}
\newcommand{\AFFosaka}{\affiliation{Department of Physics, Osaka University, Toyonaka, Osaka 560-0043, Japan}}
\newcommand{\AFFox}{\affiliation{Department of Physics, Oxford University, Oxford, OX1 3PU, United Kingdom}}
\newcommand{\AFFqmul}{\affiliation{School of Physics and Astronomy, Queen Mary University of London, London, E1 4NS, United Kingdom}}
\newcommand{\AFFregina}{\affiliation{Department of Physics, University of Regina, 3737 Wascana Parkway, Regina, SK, S4SOA2, Canada}}
\newcommand{\AFFseoul}{\affiliation{Department of Physics, Seoul National University, Seoul 151-742, Korea}}
\newcommand{\AFFsheff}{\affiliation{Department of Physics and Astronomy, University of Sheffield, S3 7RH, Sheffield, United Kingdom}}
\newcommand{\AFFshizuokasc}{\affiliation{Department of Informatics in
Social Welfare, Shizuoka University of Welfare, Yaizu, Shizuoka, 425-8611, Japan}}
\newcommand{\AFFstfc}{\affiliation{STFC, Rutherford Appleton Laboratory, Harwell Oxford, and Daresbury Laboratory, Warrington, OX11 0QX, United Kingdom}}
\newcommand{\AFFskk}{\affiliation{Department of Physics, Sungkyunkwan University, Suwon 440-746, Korea}}
\newcommand{\AFFtodai}{\affiliation{Department of Physics, University of Tokyo, Bunkyo, Tokyo 113-0033, Japan }}
\newcommand{\AFFtit}{\affiliation{Department of Physics,Tokyo Institute of Technology, Meguro, Tokyo 152-8551, Japan }}
\newcommand{\AFFtus}{\affiliation{Department of Physics, Faculty of Science and Technology, Tokyo University of Science, Noda, Chiba 278-8510, Japan }}
\newcommand{\AFFtoronto}{\affiliation{Department of Physics, University of Toronto, ON, M5S 1A7, Canada }}
\newcommand{\AFFtriumf}{\affiliation{TRIUMF, 4004 Wesbrook Mall, Vancouver, BC, V6T2A3, Canada }}
\newcommand{\AFFtokai}{\affiliation{Department of Physics, Tokai University, Hiratsuka, Kanagawa 259-1292, Japan}}
\newcommand{\AFFtsinghua}{\affiliation{Department of Engineering Physics, Tsinghua University, Beijing, 100084, China}}
\newcommand{\AFFynu}{\affiliation{Department of Physics, Yokohama National University, Yokohama, Kanagawa, 240-8501, Japan}}
\newcommand{\AFFllr}{\affiliation{Ecole Polytechnique, IN2P3-CNRS, Laboratoire Leprince-Ringuet, F-91120 Palaiseau, France }}
\newcommand{\AFFbari}{\affiliation{ Dipartimento Interuniversitario di Fisica, INFN Sezione di Bari and Universit\`a e Politecnico di Bari, I-70125, Bari, Italy}}
\newcommand{\AFFnapoli}{\affiliation{Dipartimento di Fisica, INFN Sezione di Napoli and Universit\`a di Napoli, I-80126, Napoli, Italy}}
\newcommand{\AFFroma}{\affiliation{INFN Sezione di Roma and Universit\`a di Roma ``La Sapienza'', I-00185, Roma, Italy}}
\newcommand{\AFFpadova}{\affiliation{Dipartimento di Fisica, INFN Sezione di Padova and Universit\`a di Padova, I-35131, Padova, Italy}}
\newcommand{\AFFkeio}{\affiliation{Department of Physics, Keio University, Yokohama, Kanagawa, 223-8522, Japan}}
\newcommand{\AFFwinnipeg}{\affiliation{Department of Physics, University of Winnipeg, MB R3J 3L8, Canada }}
\newcommand{\AFFkcl}{\affiliation{Department of Physics, King's College London, London, WC2R 2LS, UK }}
\newcommand{\AFFwarwick}{\affiliation{Department of Physics, University of Warwick, Coventry, CV4 7AL, UK }}
\newcommand{\AFFral}{\affiliation{Rutherford Appleton Laboratory, Harwell, Oxford, OX11 0QX, UK }}
\newcommand{\AFFwu}{\affiliation{Faculty of Physics, University of Warsaw, Warsaw, 02-093, Poland }}
\newcommand{\AFFbcit}{\affiliation{Department of Physics, British Columbia Institute of Technology, Burnaby, BC, V5G 3H2, Canada }}
\newcommand{\AFFtohoku}{\affiliation{Department of Physics, Faculty of Science, Tohoku University, Sendai, Miyagi, 980-8578, Japan }}
\newcommand{\AFFicise}{\affiliation{Institute For Interdisciplinary Research in Science and Education, ICISE, Quy Nhon, 55121, Vietnam }}
\newcommand{\AFFilance}{\affiliation{ILANCE, CNRS - University of Tokyo International Research Laboratory, Kashiwa, Chiba 277-8582, Japan}}
\newcommand{\AFFibs}{\affiliation{Institute for Basic Science (IBS), Daejeon, 34126, Korea}}

\AFFicrr
\AFFkashiwa
\AFFicrronly
\AFFmad
\AFFbu
\AFFbcit
\AFFuci
\AFFcsu
\AFFcnm
\AFFduke
\AFFllr
\AFFfukuoka
\AFFgifu
\AFFgist
\AFFuh
\AFFibs
\AFFicise
\AFFicl
\AFFbari
\AFFnapoli
\AFFpadova
\AFFroma
\AFFilance
\AFFkeio
\AFFkek
\AFFkcl
\AFFkobe
\AFFkyoto
\AFFliv
\AFFmiyagi
\AFFnagoya
\AFFkmi
\AFFpol
\AFFsuny
\AFFokayama
\AFFox
\AFFral
\AFFseoul
\AFFsheff
\AFFshizuokasc
\AFFstfc
\AFFskk
\AFFtohoku
\AFFtokai
%\AFFtokyo
\AFFtodai
\AFFipmu
\AFFtit
\AFFtus
\AFFtoronto
\AFFtriumf
\AFFtsinghua
\AFFwu
\AFFwarwick
\AFFwinnipeg
\AFFynu

%%%%%%%%%%%%%%%%%%%%%%%%%%%%%%%%%%%%%%%%%%%%%%%%%%%%%%%%%%%%%%%%%%%%
%First Author
\author[0000-0003-3273-946X]{M.~Harada}
\AFFokayama
%%%%%%%%%%%%%%%%%%%%%%%%%%%%%%%%%%%%%%%%%%%%%%%%%%%%%%%%%%%%%%%%%%%%
%ICRR
\author{K.~Abe}
\AFFicrr
\AFFipmu
\author[0000-0001-9555-6033]{C.~Bronner}
\AFFicrr
\author[0000-0002-8683-5038]{Y.~Hayato}
\AFFicrr
\AFFipmu
\author[0000-0003-1229-9452]{K.~Hiraide}
\AFFicrr
\AFFipmu
\author[0000-0002-8766-3629]{K.~Hosokawa}
\AFFicrr
\author[0000-0002-7791-5044]{K.~Ieki}
\author[0000-0002-4177-5828]{M.~Ikeda}
\AFFicrr
\AFFipmu
\author{J.~Kameda}
\AFFicrr
\AFFipmu
\author{Y.~Kanemura}
\author{R.~Kaneshima}
\author{Y.~Kashiwagi}
\AFFicrr
% ...

%%%%%%%%%%%%%%%%%%%%%%%%%%%%%%%%%%%%%%%%%%%%%%%%%%%%%%%%%%%%%%%%%%%%%

\collaboration{242}{The Super-Kamiokande Collaboration}
\noaffiliation

%\input{authors-20221203}

%\author[0000-0002-0786-7307]{Greg J. Schwarz}
%\affiliation{American Astronomical Society \\
%1667 K Street NW, Suite 800 \\
%Washington, DC 20006, USA}

%\author{August Muench}
%\affiliation{American Astronomical Society \\
%1667 K Street NW, Suite 800 \\
%Washington, DC 20006, USA}

%\collaboration{20}{(AAS Journals Data Editors)}

%\author{F.X Timmes}
%\affiliation{Arizona State University}
%\affiliation{AAS Journals Associate Editor-in-Chief}

%\author{Amy Hendrickson}
%\altaffiliation{AASTeX v6+ programmer}
%\affiliation{TeXnology Inc.}

%\author{Julie Steffen}
%\affiliation{AAS Director of Publishing}
%\affiliation{American Astronomical Society \\
%1667 K Street NW, Suite 800 \\
%Washington, DC 20006, USA}

%% Note that the \and command from previous versions of AASTeX is now
%% depreciated in this version as it is no longer necessary. AASTeX 
%% automatically takes care of all commas and "and"s between authors names.

%% AASTeX 6.31 has the new \collaboration and \nocollaboration commands to
%% provide the collaboration status of a group of authors. These commands 
%% can be used either before or after the list of corresponding authors. The
%% argument for \collaboration is the collaboration identifier. Authors are
%% encouraged to surround collaboration identifiers with ()s. The 
%% \nocollaboration command takes no argument and exists to indicate that
%% the nearby authors are not part of surrounding collaborations.

%% Mark off the abstract in the ``abstract'' environment. 
\begin{abstract}
We report the first search result for the flux of astrophysical electron antineutrinos for energies $\mathcal{O}(10)~\rm MeV$ in the gadolinium-loaded Super-Kamiokande (SK) detector.
In June 2020, gadolinium was introduced to the ultra-pure water of the SK detector in order to detect neutrons more efficiently. In this new experimental phase, SK-Gd, we can search for electron antineutrinos via inverse beta decay with efficient background rejection thanks to the high efficiency of the neutron tagging technique.
In this paper, we report the result for the initial stage of SK-Gd, during August 26th, 2020, and June 1st, 2022, with a $22.5\times 552$ kton$\cdot$day exposure at 0.01\% Gd mass concentration.
No significant excess over the expected background in the observed events is found for the neutrino energies below 31.3 MeV. Thus, the flux upper limits are placed at the 90\% confidence level.
The limits and sensitivities are already comparable with the previous SK result  with pure water ($22.5\times 2970$ kton$\cdot$day) owing to the enhanced neutron tagging.
Operation with Gd increased to 0.03\% started in June 2022.
%The operation with increased Gd to be 0.03\%, starting in June 2022, is mentioned for future prospects.
\end{abstract}

%% Keywords should appear after the \end{abstract} command. 
%% The AAS Journals now uses Unified Astronomy Thesaurus concepts:
%% https://astrothesaurus.org
%% You will be asked to selected these concepts during the submission process
%% but this old "keyword" functionality is maintained in case authors want
%% to include these concepts in their preprints.
%\keywords{Classical Novae (251) --- Ultraviolet astronomy(1736) --- History of astronomy(1868) --- Interdisciplinary astronomy(804)}

%% From the front matter, we move on to the body of the paper.
%% Sections are demarcated by \section and \subsection, respectively.
%% Observe the use of the LaTeX \label
%% command after the \subsection to give a symbolic KEY to the
%% subsection for cross-referencing in a \ref command.
%% You can use LaTeX's \ref and \label commands to keep track of
%% cross-references to sections, equations, tables, and figures.
%% That way, if you change the order of any elements, LaTeX will
%% automatically renumber them.
%%
%% We recommend that authors also use the natbib \citep
%% and \citet commands to identify citations.  The citations are
%% tied to the reference list via symbolic KEYs. The KEY corresponds
%% to the KEY in the \bibitem in the reference list below. 

\section{Introduction} \label{sec:intro}
Astrophysical electron antineutrinos ($\bar{\nu}_e$) are a unique probe to assess various physical phenomena in the universe, such as past supernovae~\citep{Beacom2010}, resonant spin flavor precession (RSFP) of solar neutrinos \citep{Akhmedov2003, Diaz2009}, and the annihilation of MeV-scale light dark matter \citep{Palomares2008}, which are expected to appear at energies $\mathcal{O}(10)~\rm MeV$. 
%Although neutrino detectors worldwide have been searching for these signals, no significant signal has been found so far.

The core-collapse supernova (CCSN) is one of the transient processes with the highest neutrino production in our universe.
A neutrino observation from CCSN has occurred only once so far, from the supernova SN1987A, by the Kamiokande \citep{Hirata1987}, IMB \citep{Bionta1987}, and Baksan \citep{Alekseev1987} neutrino detectors.
Although nearby CCSNe are rare, neutrinos from all past supernovae (Diffuse Supernova Neutrino Background; DSNB) should exist around us.
The theoretical expectations for the DSNB flux depend on various parameters: the supernova rate introduced from the cosmic star formation rate depending on the redshift, the neutrino mass ordering, the equations of states for remnant neutron stars, the metallicity of the galaxy, the failed supernova rate, and the binary interaction effect of stars \citep{Malaney1997, Hartmann1997, Kaplinghat2000, Ando2003, Ando2005, Lunardini2009, Horiuchi2009, Galais2010, Nakazato2015, Horiuchi2018, Kresse2021, Tabrizi2021, Horiuchi2021, Nick2022}. 
%The normalization of DSNB flux is determined by not only the supernova rate introduced from the cosmic star formation rate depending on the redshift, but also more various parameters, such as the neutrino mass ordering, the equation of states for remnant neutron stars, the metallicity of the galaxy, failed supernova rate, and the binary interaction effect of stars, as proposed by the theoretical expectations \citep{Malaney1997, Hartmann1997, Kaplinghat2000, Ando2003, Ando2005, Lunardini2009, Horiuchi2009, Galais2010, Nakazato2015, Horiuchi2018, Kresse2021, Tabrizi2021, Horiuchi2021, Nick2022}. 
\citet{Ashida2022} have investigated the fraction for the failed CCSNe forming black holes from the DSNB flux upper limit.

The sun is one of the most intense astrophysical electron neutrino sources at Earth. The solar $\bar{\nu}_e$ can be produced by the combination of the Mikheyev-Smirnov-Wolfenstein (MSW) effect \citep{Smirnov2005} and the RSFP effect via the neutrino magnetic moment. The limit for the neutrino-antineutrino conversion probability $P_{\nu_e\to\bar{\nu}_e}$ of solar neutrinos due to RSFP was placed from the $\bar{\nu}_e$ flux upper limit by \citet{KamLAND2022, Abe2022b, Agostini2021}. In addition, a small flux of the $\bar{\nu}_e$ from the sun, via the beta decay of $\rm{}^{40}K$, $\rm{}^{238}U$, and $\rm{}^{232}Th$, are predicted by~\citet{Malaney1990}, but also have not been observed yet.

Neutrino production from self-annihilation of MeV-scale light-dark-matter particles ($\chi\chi\to\nu\nu$) is also predicted~\citep{Arguelles2021}.
An upper limit on the averaged cross-section for the self-annihilation of the light-dark matter is placed from the $\bar{\nu}_e$ flux upper limit by \citet{KamLAND2022}.

Although neutrino detectors worldwide have been searching for these events, no significant signal has been found so far. 
The most stringent upper limits for the astrophysical $\bar{\nu}_e$ flux are set by Super-Kamiokande above 13.3~MeV \citep{Abe2021} and by KamLAND experiment below 13.3~MeV \citep{KamLAND2022}. For further lower energy regions, the Borexino experiment \citep{Agostini2021} gave the lowest upper limit below 8.3~MeV. 
In this paper, we present a new search for astrophysical $\bar{\nu}_e$ in the neutrino energy range of 9.3 MeV to 31.3 MeV based on the observation made with the new Super-Kamiokande detector configuration.

\section{The Super-Kamiokande Experiment} \label{sec:superk}
Super-Kamiokande (SK;~\citet{Fukuda2003}) is a water-Cherenkov detector experiment located 1000 m underground in Kamioka, Japan.
The detector is of cylindrical shape inside a tank with a diameter of 39.3 m and a height of 41.4 m. It is currently filled with 50 kilotonnes of gadolinium-doped ultra-pure water \citep{ABE2022a} as a target.
%It is optically separated into two concentric cylindrical tanks, an inner detector (ID) and an outer detector (OD). 
It is optically separated into a main cylindrical volume (Inner Detector; ID), surrounded by the Outer Detector (OD) that extends up to the inner surface of the SK tank. 
SK observes signals of neutrino interactions via the detection of Cherenkov light produced by charged particles within the ID using the ID PMTs. The ID is 33.8 m in diameter and 36.2 m in height, with 11129 20-inch photomultiplier tubes (PMTs) mounted pointing inwards to the inside of the ID tank. SK is sensitive to neutrinos with an energy range from several MeV to above 1 TeV. 
%On the other hand, 

%The OD surrounds the ID.
The OD is concentrically placed outside the ID; it is about two meters wide. It is instrumented with 1885 8-inch PMTs that are mounted on the outer side of the ID structure pointing outwards.  
The OD surface is covered by white Tyvek sheet to enhance the light reflection in the OD. The OD is primarily utilized to veto cosmic-ray muons.% and absorb backgorunds.

Since the start of SK operation in 1996, several upgrades to the detector have been carried out. 
Notably, the electronics were upgraded in 2008 \citep{Yamada2010} allowing the implementation of a new event trigger, Super High Energy (SHE), to identify events above ${\sim}6~{\rm MeV}$. In addition, the upgrade enables us to record all PMT hits within $535~\rm\mu s$ after the SHE-triggered event.
%Notably, in 2008, a new event trigger, Super High Energy (SHE) trigger with delayed hit information, was implemented to identify the events above ${\sim}6~{\rm MeV}$. 
%That had been possible due to the electronics upgrade \citep{Yamada2010}, which enabled us to record all PMT hits within $535~\rm\mu s$ after the SHE-triggered event.
The period for which the SHE trigger was implemented is called `SK-IV', and it operated until 2018.
%Since the start of SK operation in 1996, several upgrades to the detector have been carried out. Notably, in 2008, a new event trigger, Super High Energy (SHE), was implemented to identify the events above ${\sim}6~{\rm MeV}$.
%In addition, the electronics upgrade \citep{Yamada2010} enables us to record all PMT hits within $535~\rm\mu s$ after the SHE-triggered event. 
%The resulting event, which is called AFT trigger event, enables the search for accompanying neutrons.

In general, the triggers are issued using the number of ID or OD PMT hits within a 200 ns window. For the SHE trigger, the threshold is typically 60 ID PMT hits. The threshold for the OD trigger is typically 22 OD PMT hits.

Another important upgrade was the loading of Gd into the ultra-pure water of SK in order to improve neutron-tagging efficiency due to Gd's high neutron-capture cross-section and enhanced neutron-capture signal. 
%Thanks to this effect, the detection sensitivity Inverse Beta Decay (IBD; $\bar{\nu}_e+p\to e^++n$) for $\bar{\nu}_e$ is improved.
The original idea was proposed by \citet{Beacom2004}, and the loading was carried out starting in July 2020.
After dissolving gadolinium-sulfate-octahydrate $\rm Gd_2(SO_4)_3\cdot8H_2O$ in the ultra-pure water~\citep{ABE2022a}, the SK experiment started a new phase called SK-Gd.
The initial running period of SK-Gd is called `SK-VI'; it operated with 0.011\% Gd mass concentration until June 2022. Under this condition, about 50\% of neutrons are captured by the Gd with a typical time constant of $115~\rm\mu s$. 
Thermal neutron capture on Gd results in multiple gamma-ray emissions with a total energy of about 8 MeV, which can be easily distinguished from the random backgrounds from natural radioactivity and PMT dark noise. 
In this paper, 552.2 days of live time taken during SK-VI from August 2020 to June 2022 is used for the $\bar{\nu}_e$ signal search.

\section{Event Selection} \label{sec:event}
%In this analysis, IBD signals from the astrophysical neutrinos are investigated.
In this analysis, Inverse Beta Decay (IBD; $\bar{\nu}_e+p\to e^++n$) signals from the astrophysical neutrinos are investigated.
The signals consist of a positron-like event (prompt event) and a subsequent delayed neutron (delayed signal).
We search for IBD signals in the data with the SHE-triggered and successfully recorded 535~$\mu$s event from the signal energy region of 7.5--29.5~MeV for the reconstructed kinetic energy of prompt events ($E_{\rm rec}$).
% without OD trigger issued for removing muons.
In addition, the condition that the OD trigger is not issued is also required in order to remove incoming cosmic-ray muon events.

The event selection and reconstruction follow the previous SK-IV search \citep{Abe2021}, except for the delayed neutron identification. 
Prompt events with exactly one delayed-tagged neutron signal are selected as IBD signal candidates. 
%The prompt events with a delayed-tagged neutron signal equal to one are selected as the IBD signal candidate. 
Those neutron signals are searched for within a $535~\rm\mu s$ window after the prompt event. In SK-Gd, the neutron signal from thermal neutron capture on the Gd nucleus is efficiently tagged by a simple signal selection without using the Machine-Learning based cut used in previous analysis \citep{Abe2021}.

As a primary selection, an algorithm searching for PMT hit clusters within 200 ns windows is applied.
In this stage, the clusters with 25 hits or more are selected as candidates for neutron signals.
Then, basic event reduction cuts, used in common with the analysis for the $\mathcal{O}(10)~\rm MeV$ scale in SK, are applied. The cuts include the goodness of reconstruction, the electron likeness of the Cherenkov ring hit pattern, and the fiducial volume cut that requires the event to be further than 2 m from the wall.

For further reduction, we apply a `distance' cut and an `energy' cut.
The reconstructed vertex of the delayed signal tends to be close to the prompt event for the IBD interactions of $\mathcal{O}(10)~\rm MeV$ neutrinos.
Thus the distance between the prompt and delayed signal candidate is an efficient discriminator to select neutron events. In this analysis, the candidates that have a distance of over 3 m from the prompt event vertex are eliminated.
%The vertex of the delayed signal is reconstructed relatively close to the prompt event vertex, as shown in the middle right panel of Figure 4 in \citet{Harada2022}. Thus the distance between the prompt and delayed signal candidate is an efficient discriminator to select neutron event. In this analysis, the candidates that have  a distance of over 3 m are eliminated.
Also, the reconstructed energy of multiple gamma-ray events from Gd capture is typically around $4$--$5$~MeV, as shown in \citet{Harada2022}.
After the 3 m distance cut, the remaining background events are low-energy accidental backgrounds due to noise hits. Hence a 3.5 MeV energy threshold is assigned in order to reduce the accidental coincidence rate to $\mathcal{O}(10^{-4})~{\rm per~event}$.

Neutron tagging efficiency and the misidentification probability are estimated by applying the above selection procedure to neutron signals simulated by a Monte Carlo (MC) simulation based on Geant4.10.5p1 \citep{Allison2006}, and SK random trigger data, respectively. The Gd concentration in the MC simulation is tuned to reproduce the time constant of the measured data.
For the gamma-ray emission from thermal neutron capture on Gd in MC simulation, the ANNRI-Gd model \citep{Hagiwara2019, Tanaka2020} is utilized.
The selection efficiency of the Gd capture signal is evaluated to be $73.2~\pm~0.2\%$. 
%The selection efficiency of the Gd and the hydrogen capture signal are evaluated to be $73.2\pm0.2\%$ and $0.6\pm0.1\%$. 
Given that the capture fraction on Gd is $47.8\pm0.2\%$ and systematic uncertainty estimated based on \citet{Harada2022}, the total neutron tagging efficiency is estimated to be $35.6\pm2.5\%$.
%Given that the capture fraction on Gd is $47.8\pm0.2\%$ and systematic uncertainty estimated based on \citet{Harada2022}, $35.6\pm2.5\%$ of neutrons can be tagged.
%Considering about 50\% of the capture fraction on Gd, 39.6\% of neutrons can be tagged.
%In contrast, 
The misidentification probability of noise candidates $\varepsilon_{\rm mis}$ is found to be $(2.8\pm0.1)\times10^{-4}$~per~event.
It is sufficiently low to remove most accidental coincidences. In the previous analysis \citep{Abe2021}, typical neutron tagging efficiency and the misidentification probability using a boosted decision tree \citep{Chen2016} were about 20\% and also $\mathcal{O}(10^{-4})$, respectively.

\begin{figure}[ht!]
\centering
\includegraphics[width=\linewidth]{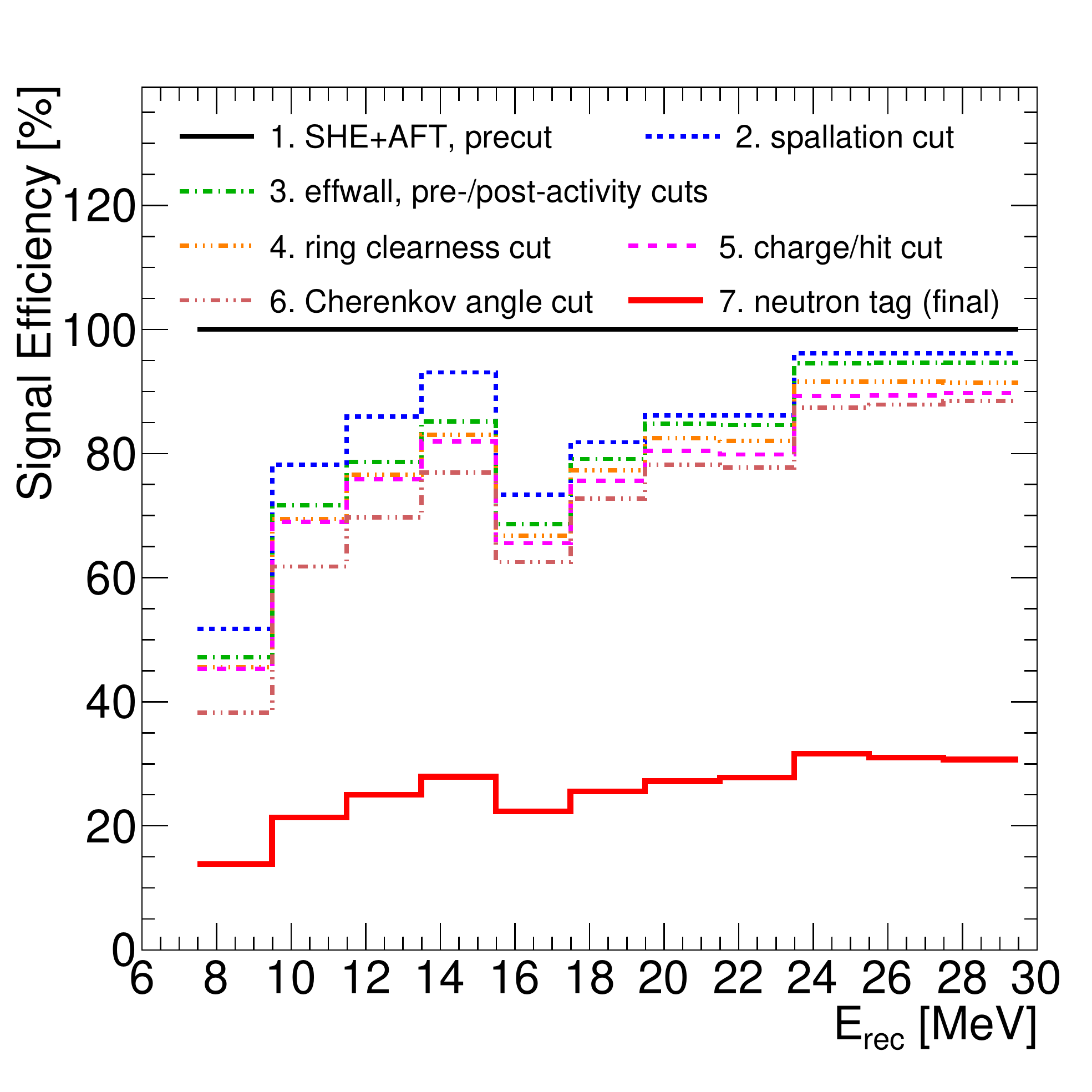}
\caption{IBD signal efficiency for the signal energy region. The 100\% efficiency line corresponds to data after trigger requirements and noise reduction cuts. These lines show the cumulative efficiency at each cut stage, performed in the order shown in the legend.
%\caption{IBD signal efficiency for the signal energy region. It started from events after the trigger requirements and the noise reduction cuts.
More detailed explanations for each reduction step, except for the neutron tagging, are described in \citet{Abe2021}. 
\label{fig:sigeff}}
\end{figure}

Figure \ref{fig:sigeff} shows the total IBD signal efficiency for each 2 MeV $E_{\rm rec}$ bin. The efficiencies after each of the main selection criteria applied are shown together.

The signal efficiency after neutron tagging (the red line in Figure~\ref{fig:sigeff}) is about twice that of the previous SK search (shown in the upper panel of Figure 18 in \citet{Abe2021}), especially below 15.49~MeV, thanks to the higher neutron detection efficiency. 

A large part of the accidental coincidence background originates from pairs of events from the decay of isotopes produced by muon spallation induced by the cosmic-ray muon and the low-energy noise events.
Therefore, efficient spallation event reduction and low $\varepsilon_{\rm mis}$ in neutron tagging are important to reduce the accidental coincidence background.
Thus, the spallation cut criteria are optimized in each energy range to remove various isotope decays at the corresponding energy.
Because of the lower $\varepsilon_{\rm mis}$ than for SK-IV, we re-optimized the spallation cut criteria below 15.49 MeV, where we have large enough statistics of the spallation sample. As a result, the spallation cut condition was loosened to achieve higher signal efficiency. On the other hand, the spallation cut criteria can not be optimized above 15.49 MeV because there are fewer spallation samples. In order to avoid possible systematics due to misestimating the spallation backgrounds in this region, harder spallation cut with lower signal efficiency was applied to almost completely eliminate spallation events above 15.49 MeV.

The side-band region above 29.49 MeV is used as the reference for the atmospheric neutrino events. The efficiency in this region is stable and almost the same as for the 27.49--29.49~MeV bin.

%The signal efficiency after neutron tagging (the red line in Figure~\ref{fig:sigeff}) is about twice that of the previous SK search (the brown line in Figure~\ref{fig:sigeff}), especially below 17.5~MeV bin.
%The signal efficiency after neutron tagging (the red line in Figure \ref{fig:sigeff}) is about twice that of the previous SK search (shown in the upper panel of Figure 18 in \citet{Abe2021}) thanks to the higher neutron detection efficiency.

\section{Background Estimation} \label{subsec:background}
Major background sources in the signal energy region are atmospheric neutrino interaction events, reactor neutrino events, and decays of spallation isotopes produced by cosmic-ray muon events.
The solar neutrinos, which predominantly interact via electron scattering in SK, have no subsequent neutron signal. Thus, requiring one neutron makes the solar neutrino background negligible.
Another background source is accidental background events which are a pair of prompt SHE events and a misidentified delayed signal due to PMT noise hits or radioactivity.

Spallation isotope decays other than the $\beta+n$ decay are efficiently removed by the spallation event cut and neutron tagging. ${}^8 \rm He$, ${}^{11} \rm Li$, ${}^{16} \rm C$, and ${}^9 \rm Li$ are the representative decay isotopes that undergo $\beta+n$ decay.
However, ${}^{11} \rm Li$ is negligible since it can be removed by the spallation cut efficiently due to its short life ($T_{1/2} < 0.01~\rm s$), and its production yield is low ($10^{-9}~\rm \mu^{-1}g^{-1}cm^2$) \citep{Li2014}.
Also, ${}^8 \rm He$ and ${}^{16} \rm C$ have a low production yield of about $0.23\times 0.16$ and $0.02\times 10^{-7}~\rm \mu^{-1}g^{-1}cm^2$ \citep{Li2014}.
In contrast, $\rm {}^9Li$ has a relatively long life ($T_{1/2}\sim 0.18~\rm s$), so it is difficult to identify its parent muon.
In addition, the production yield is sufficiently higher, about $1.9\times 0.51 \times 10^{-7}~\rm \mu^{-1}g^{-1}cm^2$ \citep{Li2014}, where 0.51 comes from the branching ratio for this decay mode. 
Thus, we consider only $\rm {}^9Li$ as the remaining background from the spallation isotope decays. 
The production yield of $\rm {}^9Li$ event is measured to be $0.86 \pm 0.12 \, (\mathrm{stat.}) \pm 0.15 \, (\mathrm{sys.})$~${\rm kton^{-1}\cdot day^{-1}}$ by \citet{Zhang2016}, and a prediction of the spectrum is given in \citet{Abe2021}.

Reactor $\bar{\nu}_e$ inevitably remains in the final signal candidates since the observed event is IBD which is the same as our target signal. The spectrum and yield are evaluated by the SKReact code \citep{SKReact} based on the reactor neutrino model from \citet{Baldoncini2015} in the first 7.5--9.5 MeV $E_{\rm rec}$ bin. Their contribution above the 9.5 MeV $E_{\rm rec}$ bin is negligible. 

The number of atmospheric neutrino background events is estimated by the simulation based on the HKKM flux \citep{Honda2007, Honda2011} as the neutrino flux and NEUT 5.4.0.1 \citep{Hayato2009, Hayato2021} as the neutrino interaction simulator. 
Below 16 MeV of $E_{\rm rec}$, nuclear deexcitation gamma-rays from neutral-current quasi-elastic (NCQE) interactions dominate \citep{Ankowski2012}. NCQE interactions induce prompt gamma rays and nuclei production. Above 16~MeV, it is dominated by charged-current quasi-elastic (CCQE) interactions with neutron production.
Most of the events with $E_{\rm rec}$ from 29.5 to 79.5 MeV consist of the decay electron from an unobserved muon originating from CCQE interacting atmospheric neutrinos.
Since its energy distribution is the well-known Michel spectrum, the flux of atmospheric neutrino CCQE events in our MC is scaled by fitting the spectrum of the $29.5 < E_{\rm rec} < 79.5 ~\rm MeV$ side-band region to the data.

The number of accidental coincidence background events $B_{\rm acc}$ is estimated as 
\begin{equation}
    B_{\rm acc} = \varepsilon_{\rm mis} \times N_{\rm pre\mathchar`-ntag}^{\rm data},
\end{equation}
where $ \varepsilon_{\rm mis}$ is the neutron misidentification probability described in Section \ref{sec:event}, and $N_{\rm pre\mathchar`-ntag}^{\rm data}$ represents the number of remaining observed events after all selection criteria except neutron tagging.
 
Systematic uncertainties are estimated for only signal energy regions.
The uncertainties on the NCQE events, spallation ${}^9\rm Li$, and reactor neutrinos are taken as estimated by \citet{Abe2021}, as the 68\% below 15.49~MeV and 82\% above 15.49~MeV, 60\%, and 100\% for the NCQE, ${}^9\rm Li$, and reactor neutrino backgrounds, respectively. Other components, such as non-NCQE events and accidental coincidence events, are newly estimated from the observed data in SK-VI based on the same method as \citet{Abe2021}, 44\% and 4\%, respectively.

\section{Results} \label{subsec:result}
\begin{figure}[ht!]
\begin{center}
\includegraphics[width=\linewidth]{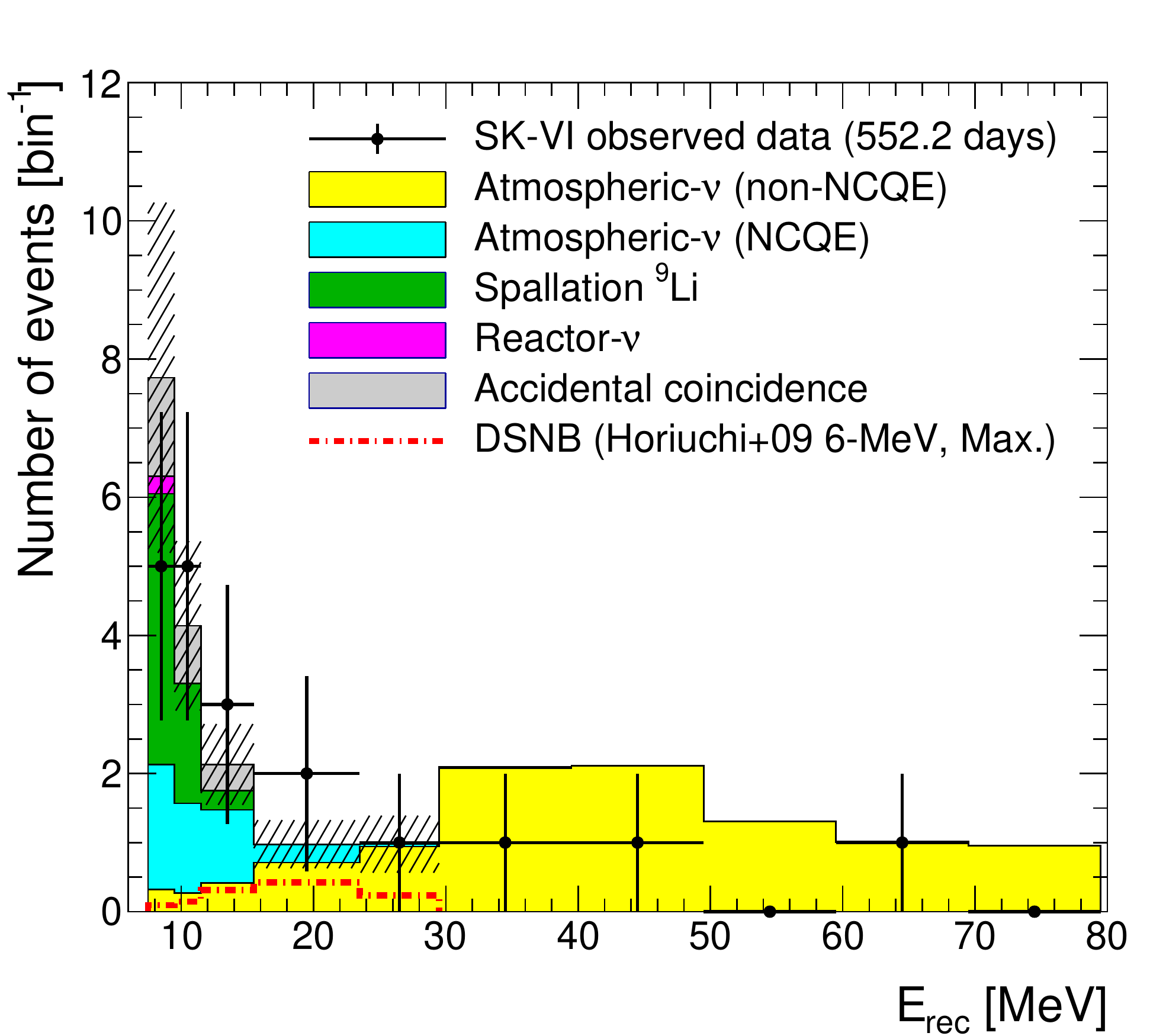}

\includegraphics[width=\linewidth]{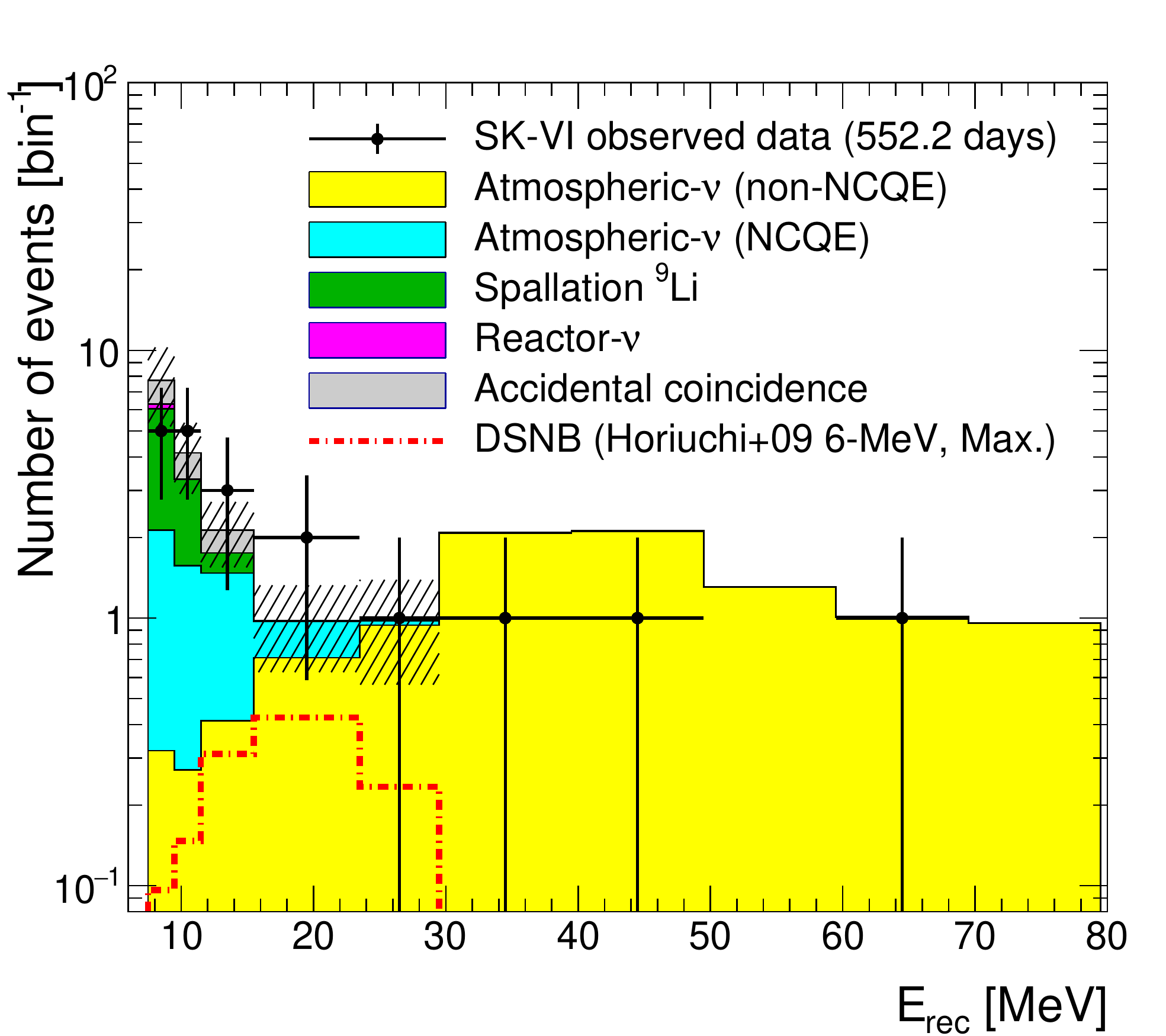}
\end{center}
\caption{Reconstructed energy spectra of the observed data and the expected background after data reductions with a linear (top) and a logarithmic (bottom) scale for the vertical axis. These include the signal energy region and the side-band region above 29.5 MeV. Each color-filled histogram shows the expected backgrounds. The error bars in the data points represent the statistical uncertainty estimated by taking the square root of the number of observed events. These background histograms are stacked on the other histograms. The hatched areas represent the total systematic uncertainty for each bin. The size of uncertainty for each background is mentioned in the main text. The red dot-dashed line shows the DSNB expectation from the Horiuchi+09 model~\citep{Horiuchi2009}, which is drawn separately from the stacked histogram of the estimated backgrounds.
\label{fig:spectrum}}
\end{figure}

After all event selection criteria are applied, 16 events remain within the signal energy region in 552.2~days data.
In this analysis, we adopt five separate bins of $E_{\rm rec}$, of widths 7.5--9.5, 9.5--11.5, 11.5--15.5, 15.5--23.5, and 23.5--29.5~MeV for the signal window. Also, the side-band region is separated into bins for each 10~MeV. Figure \ref{fig:spectrum} shows the $E_{\rm rec}$ spectrum of those events.
This is also listed in Table \ref{tab:eventsum}.

The probabilities of finding the observed number of events due to the fluctuation of the background events ($p$-value) are evaluated for each bin.
It is done by performing $10^6$ pseudo experiments based on the number of observed events and expected background events and the systematic uncertainties of the latter. The obtained $p$-values are listed in Table \ref{tab:eventsum}.
%Table \ref{tab:eventsum} summarizes the observed event, expected background event, and $p$-values for each bin. 
We conclude that no significant excess is observed in the data over the expected background since even the most significant bin has a $p$-value is 25.8\%.

\begin{deluxetable}{c c c c}[h!]
\tablecaption{Summary of observed events, expected background events, and $p$-value for each $E_{\rm rec} $ bin. Errors for the expected background represent only the systematic uncertainty. \label{tab:eventsum}} 
\tablecolumns{4}
\tablewidth{0pt}
\tablehead{
    \colhead{$E_{\rm rec}$ [MeV]} & 
    \colhead{Observed} & 
    \colhead{Expected} &
    \colhead{$p$-value}
}
\startdata
7.5--9.5   & 5 & $7.73\pm 2.54$ & 0.798 \\
9.5--11.5  & 5 & $4.14\pm 1.23$ & 0.398 \\
11.5--15.5 & 3 & $2.13\pm 0.59$ & 0.359 \\
15.5--23.5 & 2 & $0.98\pm 0.35$ & 0.258 \\
23.5--29.5 & 1 & $0.98\pm 0.41$ & 0.597 \\
\enddata
\end{deluxetable}

\vspace{-10ex}
We set the upper limit for the number of signal excess over the expected background with a 90\% confidence level (C.L.) ($N_{\rm 90}$).
It is evaluated by the pseudo experiments using the number of observed events with these $1\sigma$ statistical uncertainties and the number of expected background events with their systematic uncertainties.
Then we estimate the flux upper limit based on $N_{90}$ of the observed event. 
%, as well as on the expected background (expected sensitivity). 
Assuming there is no signal event, the upper limit on the flux for each bin is calculated as
\begin{equation}
        \label{eq:sensitivity}
      \phi^{\rm limit}_{\rm 90} = \frac{N_{90}}{\bar{\sigma}_{\rm IBD} \cdot N_p \cdot T \cdot \bar{\varepsilon}_{\rm sig}\cdot dE}.
%      \phi^{\rm limit}_{\rm 90} = \frac{N_{90}}{\bar{\sigma}_{\rm IBD} \cdot N_p \cdot T \cdot \bar{\varepsilon}_{\rm sig}\cdot dE}~\rm [cm^{-2}\, ~s^{-1}\, MeV^{-1}].
\end{equation}
Here, $\bar{\sigma}_{\rm IBD}$ is the averaged total cross-section of IBD for each energy bin, $N_p$ is the number of protons as a target in the 22.5 kilotons of the fiducial volume of SK, $T$ is the live-time of observation  (552.2 days), $\bar{\varepsilon}_{\rm sig}$ is the averaged signal efficiency for each energy bin after all event selection criteria are applied as shown in Figure \ref{fig:sigeff}, and $dE$ is the bin width at each bin. The neutrino energy $E_\nu$ is calculated by $E_\nu = E_{\rm rec} + 1.8~\rm MeV$. The total cross-section is given by the calculation in \citet{Strumia2003}.

The expected upper limit from the background-only hypothesis at 90\% C.L., $N_{\rm 90, exp}$, is evaluated using the number of expected background events and their statistical uncertainty.
Then we extract the expected flux sensitivity by replacing $N_{\rm 90}$ with $N_{\rm 90, exp}$ in Equation~\ref{eq:sensitivity}.

\begin{figure*}[htb!]
    \centering
        \includegraphics[width=0.66\linewidth]{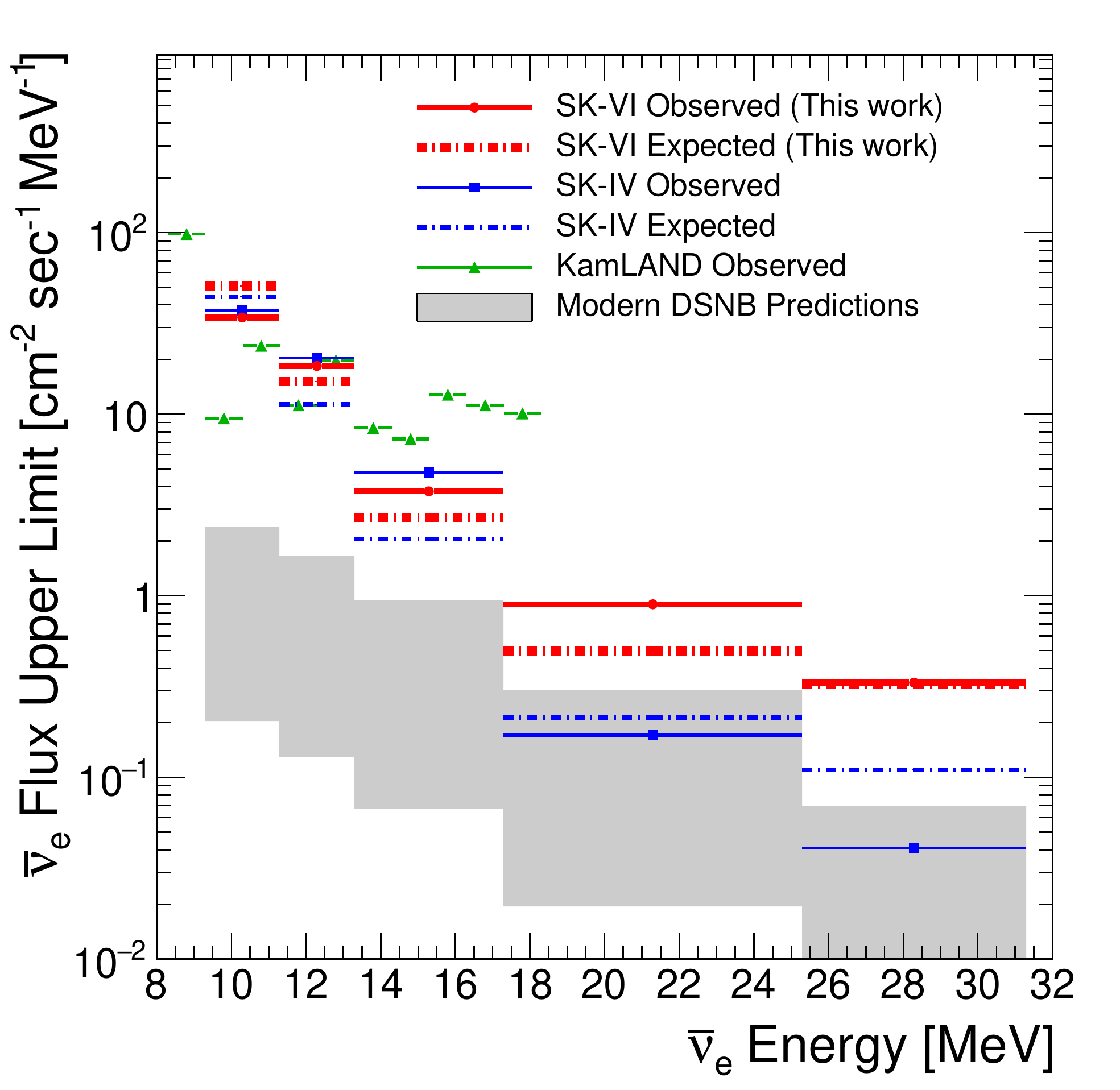}
        \caption{Upper limits on the $\bar{\nu}_e$ flux, calculated by Equation~\ref{eq:sensitivity}. The red lines show the observed (solid) and expected (dot-dashed) 90\% C.L. upper limit for SK-VI. The blue lines show the observed (solid) and expected (dot-dashed)  90\% C.L. upper limit for SK-IV~\cite{Abe2021}. The green line represents the 90\% C.L. observed upper limit placed by KamLAND~\cite{KamLAND2022}. The gray-shaded region represents the range of the modern theoretical expectation. The expectation drawn in the figure includes DSNB flux models \citep{Malaney1997, Hartmann1997, Kaplinghat2000, Ando2003, Lunardini2009, Horiuchi2009, Galais2010, Nakazato2015, Horiuchi2018, Kresse2021, Tabrizi2021, Horiuchi2021, Nick2022}. Ando+03 model was updated in \citet{Ando2005}.
%        \caption{Upper limits on the $\bar{\nu}_e$ flux. Solid lines show the observed upper limit for SK-IV (blue), SK-VI (red), and KamLAND (green), with 90\% C.L., respectively. Dot-dashed lines show the expected sensitivity based on the expected background for SK-IV (blue) and SK-VI (red). The gray-shaded region represents the range of the modern theoretical expectation. The expectation drawn in the figure includes DSNB flux models \citep{Malaney1997, Hartmann1997, Kaplinghat2000, Ando2003, Lunardini2009, Horiuchi2009, Galais2010, Nakazato2015, Horiuchi2018, Kresse2021, Tabrizi2021, Horiuchi2021, Nick2022}. Ando+03 model was updated in \citet{Ando2005}.
        \label{fig:limit}}
\end{figure*}

Figure \ref{fig:limit} shows the upper limit of the $\bar{\nu}_e$ flux extracted in this search with the range of expectations of modern DSNB models. The most optimistic expectation is Kaplinghat+00 \citep{Kaplinghat2000}, and the most pessimistic one is Nakazato+15 \citep{Nakazato2015} with the assumption of normal mass ordering in whole energy ranges, respectively.
The upper limit of the flux for each bin is summarized in Table \ref{tab:fluxsum}.

\begin{deluxetable*}{ c c c c c c}[htb!]
\centering
\tablecaption{Summary table of upper limits, sensitivity, and optimistic and pessimistic DSNB expectation from \citet{Kaplinghat2000}, and \citet{Nakazato2015}, respectively.\label{tab:fluxsum}} 
\tablecolumns{6}
\tablewidth{0pt}
\tablehead{
\colhead{Neutrino Energy} & \multicolumn{2}{c}{Observed upper limit} & \multicolumn{2}{c}{Expected sensitivity} & \colhead{Averaged theoretical expectation of DSNB} \\
\colhead{$\rm [MeV]$} & \multicolumn{2}{c}{$\rm [cm^{-2}\, s^{-1}\, MeV^{-1}]$} & \multicolumn{2}{c}{$\rm [cm^{-2}\, s^{-1}\, MeV^{-1}]$} & \colhead{$\rm [cm^{-2}\, s^{-1}\, MeV^{-1}]$} \\
%    & \colhead{SK-IV} & \colhead{SK-VI} & \colhead{SK-IV} & \colhead{SK-VI} & \colhead{\citet{Kaplinghat2000}}
    & \colhead{SK-IV} & \colhead{SK-VI} & \colhead{SK-IV} & \colhead{SK-VI} &
}
\startdata
    9.29--11.29 & 37.30 & 34.07 & 44.35 & 50.78 & 0.20 -- 2.40 \\
    11.29--13.29 & 20.43 & 18.43 & 11.35 & 15.12 & 0.13 -- 1.66 \\
    13.29--17.29 & 4.77 & 3.76 & 2.05 & 2.71 & 0.67 -- 0.94 \\
    17.29--25.29 & 0.17 & 0.90 & 0.21 & 0.50 & 0.02 -- 0.30 \\
    25.29--31.29 & 0.04 & 0.33 & 0.11 & 0.33 & $<0.01$ -- 0.07 \\
\enddata
\end{deluxetable*}

\vspace{-4ex}
\section{Future prospects}
In June 2022, the SK-Gd experiment was upgraded to the SK-VII phase, in which additional Gd was introduced into the detector, providing a mass concentration of approximately 0.03\%.
In this phase, neutron tagging efficiency is expected to be over 55\% while having comparable $\varepsilon_{\rm mis}$ with SK-VI, leading to 1.5 times higher sensitivity for the $\bar{\nu}_e$ in the case of the same live-time as for SK-VI. 
Furthermore, more efficient noise reduction by neutron tagging will enable a lower energy threshold. 
Hence we can search in a lower energy region, which will increase signal acceptance for DSNB, solar anti-neutrinos, and light-dark matter searches.

\section{Conclusions} \label{sec:conclusion}
We searched for astrophysical $\bar{\nu}_e$, using the SK-VI data below 29.5 MeV for $E_{\rm rec}$ between August 2020 and May 2022, with 0.01\% Gd mass concentration. This is an independent data set from the previous SK-IV search \citep{Abe2021}, using the data taken with pure water. 
In this analysis, a brand-new method for tagging neutrons using the signal of neutron capture on Gd is utilized so that the efficiency of neutron tagging is twice as high while keeping a low-misidentification probability.
%In this analysis, a brand-new method for tagging neutrons using the signal of neutron capture on Gd is used so that the efficiency of neutron tagging and misidentification probability is drastically improved.
No significant excess above the expected backgrounds at greater than 90\% C.L. level for five separate energy bins between 9.3 and 31.3~MeV is found.
%No significant excess above the expected backgrounds at greater than $2\sigma$ level for five separated energy bins is found.
Thus, we placed the upper limits on the $\bar{\nu}_e$ flux for the observed upper limit and the expected sensitivity.
The sensitivity of this work is comparable to the previous SK-IV search \citep{Abe2021} with a live-time of 2970 days, which is the world's most sensitive search above 13.3 MeV, even though the live-time of 552.2 days is about five times smaller.
%Comparing to the previous SK-IV search \citep{Abe2021} with 2970 days of live-time, the sensitivity of this work using the SK-VI data is comparable even though the live-time is about five times smaller than SK-IV, which is among the world's most sensitive searches.
The result was achieved by neutron tagging with Gd signal and lowered probability of accidental coincidences thanks to the benefit of introducing Gd.

%From June 2022, the SK-Gd experiment has been upgraded to the 'SK-VII' phase by introducing additional Gd to the tank. As a result, the mass concentration is about 0.03\%.
%In June 2022, the SK-Gd experiment was upgraded to the SK-VII phase, where additional Gd was introduced into the detector, providing a mass concentration of approximately 0.03\%.
%In this phase, neutron tagging efficiency is expected to be over 55\% while having comparable $\varepsilon_{\rm mis}$ with $\rm SK\mathchar`-VI$, leading to 1.5 times higher sensitivity in the case of the same live-time with $\rm SK\mathchar`-VI$. 
%Furthermore, more efficient noise reduction by neutron tagging will enable a lower energy threshold. 
%Hence we can search the lower energy region, which will increase signal acceptance for DSNB, solar anti-neutrinos, and light-dark matter searches.

%\begin{acknowledgments}
We gratefully acknowledge the cooperation of the Kamioka Mining and Smelting Company. The Super-Kamiokande experiment has been built and operated from funding by the Japanese Ministry of Education, Culture, Sports, Science and Technology, the U.S. Department of Energy, and the U.S. National Science Foundation. Some of us have been supported by funds from the National Research Foundation of Korea (NRF-2009-0083526 and NRF 2022R1A5A1030700) funded by the Ministry of Science, ICT, the Institute for Basic Science (IBS-R016-Y2), and the Ministry of Education (2018R1D1A1B07049158, 2021R1I1A1A01042256, the Japan Society for the Promotion of Science, the National Natural Science Foundation of China under Grants No.11620101004, the Spanish Ministry of Science, Universities and Innovation (grant PID2021-124050NB-C31), the Natural Sciences and Engineering Research Council (NSERC) of Canada, the Scinet and Westgrid consortia of Compute Canada, the National Science Centre (UMO-2018/30/E/ST2/00441) and the Ministry of Education and Science (DIR/WK/2017/05), Poland, the Science and Technology Facilities Council (STFC) and GridPP, UK, the European Union's Horizon 2020 Research and Innovation Programme under the Marie Sklodowska-Curie grant agreement no.754496, H2020-MSCA-RISE-2018 JENNIFER2 grant agreement no.822070, and H2020-MSCA-RISE-2019 SK2HK grant agreement no. 872549.
%We gratefully acknowledge cooperation of the Kamioka Mining and Smelting Company. The Super-Kamiokande experiment was built and has been operated with funding from the Japanese Ministry of Education, Science, Sports and Culture, and the U.S. Department of Energy.
%\end{acknowledgments}

%% For this sample we use BibTeX plus aasjournals.bst to generate the
%% the bibliography. The sample631.bib file was populated from ADS. To
%% get the citations to show in the compiled file do the following:
%%
%% pdflatex sample631.tex
%% bibtext sample631
%% pdflatex sample631.tex
%% pdflatex sample631.tex

\bibliography{library2}{}
\bibliographystyle{aasjournal}

%% This command is needed to show the entire author+affiliation list when
%% the collaboration and author truncation commands are used.  It has to
%% go at the end of the manuscript.
%\allauthors

%% Include this line if you are using the \added, \replaced, \deleted
%% commands to see a summary list of all changes at the end of the article.
%\listofchanges

\end{document}